\documentclass[10pt,aps,prc,floatfix,twocolumn,nofootinbib, superscriptaddress,preprintnumbers]{revtex4-1}
\usepackage{amsfonts,amsmath,amssymb,bm, xspace}
\usepackage{color,graphicx}
\usepackage{bbm,bm}
\usepackage{dcolumn}                 
\usepackage[dvipsnames]{xcolor}
\usepackage{bbold}
\renewcommand{\epsilon}{\varepsilon}

\makeatletter
\g@addto@macro\bfseries{\boldmath}
\makeatother

\usepackage{hyperref}                

\usepackage{array,physics}
\usepackage{dcolumn}
\usepackage[normalem]{ulem}
\newcolumntype{d}[1]{D{.}{.}{#1}}

\hypersetup{%
    pdfsubject=Paper,
    unicode=true,
    breaklinks=true,
     colorlinks   = true,
     linkcolor = blue,
     citecolor = blue,
     menucolor = blue,
     urlcolor = blue
}

\newcommand{\sat}{\mathrm{sat}}

\newcommand{\tov}{\mathrm{TOV}}

\newcommand{\term}{\mathrm{term}}
\newcommand{\pQCD}{\mathrm{pQCD}}

\begin{document}

\preprint{LA-UR-23-32625}

\title{Equation of state at neutron-star densities and beyond from perturbative QCD}

\author{Oleg Komoltsev}
\email{oleg.komoltsev@uis.no}
\affiliation{Faculty of Science and Technology, University of Stavanger, 4036 Stavanger, Norway}

\author{Rahul Somasundaram}
\email{rsomasun@syr.edu}
\affiliation{Theoretical Division, Los Alamos National Laboratory, Los Alamos, New Mexico 87545, USA}
\affiliation{Department of Physics, Syracuse University, Syracuse, NY 13244, USA}

\author{Tyler Gorda}
\email{gorda@itp.uni-frankfurt.de}
\affiliation{Institut f\"ur Theoretische Physik, Goethe Universit\"at,   Max-von-Laue-Str. 1, 60438 Frankfurt am Main, Germany}
\affiliation{Technische Universit\"at Darmstadt, Department of Physics, 64289 Darmstadt, Germany}
\affiliation{ExtreMe Matter Institute EMMI, GSI Helmholtzzentrum f\"ur Schwerionenforschung GmbH, 64291 Darmstadt, Germany}

\author{Aleksi Kurkela}
\email{aleksi.kurkela@uis.no}
\affiliation{Faculty of Science and Technology, University of Stavanger, 4036 Stavanger, Norway}

\author{J\'{e}r\^{o}me Margueron}
\email{j.margueron@ip2i.in2p3.fr}
\affiliation{Institut de Physique des 2 infinis de Lyon, CNRS/IN2P3, Universit\'e de Lyon, Universit\'e Claude Bernard Lyon 1, F-69622 Villeurbanne Cedex, France
}
\affiliation{International Research Laboratory on Nuclear Physics and Astrophysics, Michigan State University and CNRS, East Lansing, MI 48824, USA} 

\author{Ingo Tews}
\email{itews@lanl.gov}
\affiliation{Theoretical Division, Los Alamos National Laboratory, Los Alamos, New Mexico 87545, USA}

\date{\today}
\begin{abstract}
We explore the consequences of imposing robust thermodynamic constraints arising from perturbative Quantum Chromodynamics (QCD) when inferring the dense-matter equation-of-state (EOS).
We find that the termination density, up to which the EOS modeling is performed in an inference setup, strongly affects the constraining power of the QCD input.
This sensitivity in the constraining power arises from EOSs that have a specific form, with drastic softening immediately above the termination density followed by a strong stiffening.
We also perform explicit modeling of the EOS down from perturbative-QCD densities to construct a new QCD likelihood function that incorporates additional perturbative-QCD calculations of the sound speed and is insensitive to the termination density, which we make publicly available.
\end{abstract}

\maketitle

\section{Introduction}

Studying the impact of perturbative quantum chromodynamics (pQCD) calculations on the inference of the equation of state (EOS) of neutron stars (NSs) has recently become a promising area of research. 
The fundamental theory of the strong interaction, quantum chromodynamics (QCD) is non-perturbative at low energies but becomes perturbative at asymptotically high energies, which are reached either at high temperatures or at high densities, at about $20-40 n_\sat$ in baryon number density~\cite{Kurkela:2009gj,Kurkela:2016was,Gorda:2018gpy,Gorda:2021znl,Gorda:2023mkk}.
Here, $n_\sat\approx 0.16$~fm$^{-3}$ is the nuclear saturation density.

The cores of stable NSs explore densities that are most likely limited to about $5 - 8 n_\sat$, depending on the EOS which is realized in their cores~\cite{Schaffner-Bielich:2020psc}. 
This implies that the theoretical description of dense matter in stable NSs requires the solution of QCD in a non-perturbative regime.
At low densities, up to $1-2 n_{\rm sat}$, effective-field-theory (EFT) approaches, such as chiral EFT~\cite{Epelbaum:2008ga,Machleidt:2011zz}, are commonly employed to study the NS EOS~\cite{Hebeler:2013nza,Tews:2018kmu,Drischler:2017wtt}.
Such calculations are typically extended to higher densities using model-agnostic extrapolations~\cite{Read:2008iy,Lindblom:2010bb,Lindblom:2012zi,Hebeler:2013nza,Lindblom:2013kra, Kurkela:2014vha,Annala:2017llu,Most:2018hfd,Tews:2018iwm,Landry:2018prl,Greif:2018njt,Annala:2019puf,Raaijmakers:2019dks,Essick:2020flb,Annala:2021gom,Altiparmak:2022bke,Somasundaram:2021clp,Jiang:2022tps,Gorda:2022jvk,Annala:2023cwx}, conditioned with various astrophysical inputs.
Some of these analyses have also included input from the fundamental QCD Lagrangian in the perturbative regime (henceforth referred to as ``QCD input'') in the form of pQCD calculations at high densities~\cite{Kurkela:2014vha, Annala:2017llu,Most:2018hfd,Annala:2019puf,Somasundaram:2021clp, Annala:2021gom,Han:2022rug, Jiang:2022tps, Gorda:2022jvk, Annala:2023cwx}.
These works have reported results that differ from those that do not account for this QCD input. 
In particular, the works including the QCD input have reported softening of the EOS at energy densities above $750$~MeV/fm$^3$, for which there seems to be no evidence without this input. 

The connection between the perturbative regime of QCD and NSs was first investigated in Ref.~\cite{Kurkela:2014vha}, by extending model-agnostic extrapolation functions of the EOS up to the densities where pQCD calculations become reliable and imposing the pQCD values as boundary conditions for these extrapolations. 
More recently, Komoltsev and Kurkela~\cite{Komoltsev:2021jzg} have suggested a new method to link both density regimes, allowing them to `integrate backwards', i.e., to propagate the pQCD constraints to lower densities in a completely general, analytical, and model-independent manner using only the thermodynamic potential and the conditions of causality and mechanical stability.
In practice, the method provides a necessary condition for an extrapolated EOS that has to be fulfilled at all densities below the validity range of pQCD calculations.
This condition asserts that all the points of the EOS can be connected to the pQCD values by some causal and stable interpolation.
The QCD input was shown to constrain the EOS down to densities of $\sim 2.2-2.5n_\sat$, if otherwise only causality and mechanical stability are considered beyond chiral EFT. 
Within this construction, the only uncertainty in the constraints arises from the truncation error of the pQCD computation of the EOS at perturbative densities (for a comprehensive study on the pQCD truncation errors, see Ref.~\cite{Gorda:2023usm}). 
However, at these densities, astrophysical observations of NSs are naturally more informative.

While the construction suggested in Ref.~\cite{Komoltsev:2021jzg} fully bypasses the need for modeling NSs (as the input to these constraints derives from high-energy particle physics, not from NSs), in order to interface the QCD input with these astrophysical observations several assumptions must be made. 
These include extrapolations involved in nuclear theory, assumptions involved when extracting NS properties from observational data, the prior assumptions of how EOS models are constructed, and up to what terminating number density $n_\mathrm{term}$ such modeling is performed.
These assumptions and their interplay with the pQCD calculations are an important subject in discussions about the constraining power of  astrophysical and QCD inputs.

The constraining power of pQCD calculations was studied in detail in two recent publications~\cite{Gorda:2022jvk,Somasundaram:2022ztm}, and was later also investigated by Refs.~\cite{Ecker:2022dlg,Annala:2023cwx,Essick:2023fso,Brandes:2023hma,Musolino:2023edi,Mroczek:2023zxo,Fan:2023spm,Tang:2023owf}.
These works have employed astrophysical input in the form of gravitational-wave data constraining the tidal deformability of NSs in binary mergers~\cite{TheLIGOScientific:2017qsa,LIGOScientific:2018cki,LIGOScientific:2018hze}, radio measurements of NS masses~\cite{Antoniadis:2013pzd,Cromartie:2019kug,Fonseca:2021wxt}, as well as simultaneous mass-radius measurements using X-ray observations~\cite{Steiner:2017vmg,Nattila:2017wtj,Shawn:2018,Miller:2019cac,Riley:2019yda,Miller:2021qha,Riley:2021pdl}.
They discussed whether the QCD input reduces the uncertainties of the inferred EOS when imposed on top of astrophysical inputs or whether the QCD and astrophysical inputs disfavor similar behaviors. 
In the latter case, the two inputs would corroborate one another, strengthening the conclusion that would have been obtained using only one. 

\begin{figure}[t]
\includegraphics[width = \columnwidth]{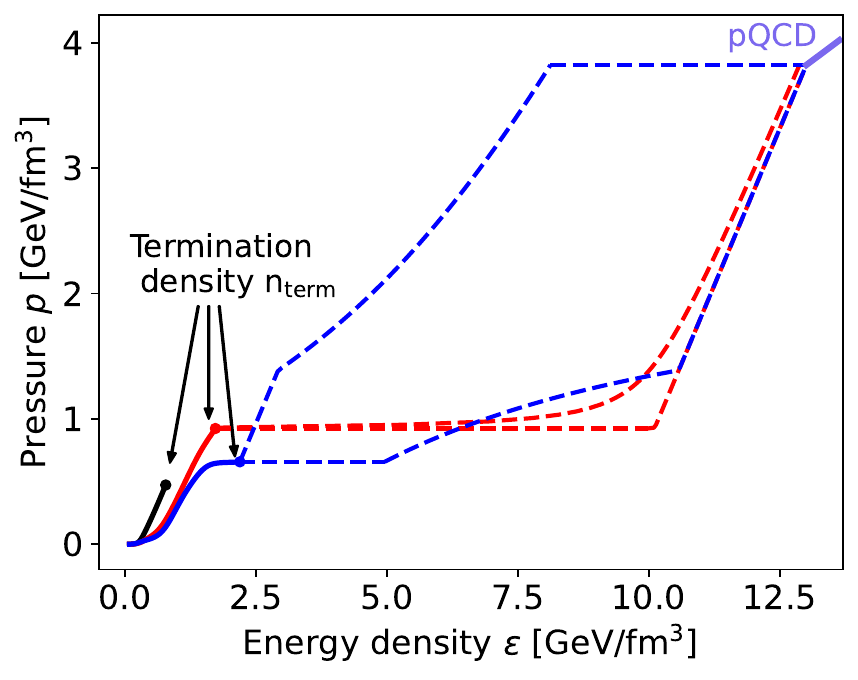}
\caption{The construction of the pQCD constraint of Ref.~\cite{Komoltsev:2021jzg}. 
The explicit form of the EOS is specified until $n = n_{\term}$. 
At this point, the pQCD constraint considers all possibilities for connecting the EOS from $n_{\term}$ to $n_{\pQCD}$, where pQCD calculations become reliable. 
The blue and red solid lines show EOSs that are compatible with pQCD at $n_\term$, having a pQCD likelihood function of 1.
However, for the blue EOS, the parameter space for the EOS at higher densities is large (indicated by the blue-dashed lines) while the EOS above $n_{\term}$ are extremely constrained for the red EOS (indicated by red-dashed lines). 
Finally, the black line corresponds to an EOS which cannot be connected to pQCD at $n_\term$ without violating causality or thermodynamic stability.
For this EOS, the pQCD likelihood function is 0.}
\label{fig:1}
\end{figure}

In this work, we compare the analyses conducted in Refs.~\cite{Gorda:2022jvk} and~\cite{Somasundaram:2022ztm}.
These studies arrived at different conclusions but used the same pQCD calculations and the same way of constraining the EOS~\cite{Komoltsev:2021jzg}, see Fig.~\ref{fig:1}.
While both of these studies agreed that the QCD input impacted the EOS at the densities reached in massive NSs, 
Ref.~\cite{Gorda:2022jvk} found that QCD input provides considerable additional information on top of other sources of data whereas Ref.~\cite{Somasundaram:2022ztm} found this additional impact to be marginal. 
The conclusions of these two studies seem to be in contradiction, and the aim of this work is to investigate the origin of this apparent discrepancy.

In addition to this direct comparison between Refs.~\cite{Gorda:2022jvk,Somasundaram:2022ztm}, we also investigate the EOS behavior beyond $n_\term$ that is required in order for the EOS to remain consistent with the QCD input to higher densities. 
We do so by generating ensembles of extensions of a given EOS beyond $n_\term$, which we condition with the QCD input at $15 n_{\rm sat}$.
In this context, we also introduce a concept of a pQCD tension index, which quantifies the degree to which a given EOS is in tension with the high-density pQCD constraint within the construction of Ref.~\cite{Komoltsev:2021jzg}, before being fully excluded.

Finally, we also introduce a method for modifying the QCD input itself by marginalizing over a prior model of high density EOSs constructed by extrapolating down from $n < n_\pQCD$. 
Such marginalization has so far only been performed within parametric models \cite{Han:2022rug, Jiang:2022tps,Annala:2023cwx} (cf.~the discussion in Ref.~\cite{Essick:2023fso}).
Importantly, such a construction also permits us to condition the model with further pQCD information not used in the original construction. 
In particular, the speed of sound exhibits very small perturbative corrections to the free result down to densities as low as $\approx 25n_\sat$. 
By conditioning the high-density prior with the speed of sound gives us a novel, more theory-informed QCD input at the price of increased model dependence.

The paper is organized as follows. 
In Sec.~\ref{sec:pQCD_review}, we review the construction of Ref.~\cite{Komoltsev:2021jzg} and define the QCD input. 
We also introduce the pQCD tension index. 
In Sec.~\ref{sec:computational_setup}, we discuss our computational setup, including the prior EOS models used below $n_\term$ and the astrophysical constraints that we use within our Bayesian analysis. 
In Sec.~\ref{sec:results}, we discuss our results.
Finally, in Sec.~\ref{sec:discussion_and_conclusions}, we discuss our findings and limitations of our work.

\section{Review of the QCD input}
\label{sec:pQCD_review}

Following Refs.~\cite{Gorda:2022jvk} and~\cite{Komoltsev:2021jzg}, we implement the QCD input as follows. 
Given the EOS at two different points, at a low density $\beta_L = (n_L, p_L, \mu_L)$ and at a high density $\beta_H =(n_H, p_H, \mu_H)$, we may ask whether these two points can be connected by \emph{any} stable and causal EOS.
Here $n$ is the baryon density, $p$ is the pressure, and $\mu$ is the baryon chemical potential.
This question can be answered by exploring the bounds for EOS behavior between $\beta_L$ and $\beta_H$. 

The range of possible pressure differences  between $(\mu_L, n_L)$ and $(\mu_H, n_H)$ are limited by the following two EOSs in this interval
\begin{align}
n_\mathrm{pmin}(\mu) &= n_L \frac{\mu}{\mu_L}, \quad \mu_L \leq \mu < \mu_H  \label{npmin}\\
n_\mathrm{pmax}(\mu) &= n_H \frac{\mu}{\mu_H}, \quad \mu_L < \mu \leq \mu_H  \label{npmax}
\end{align}
namely those which have one segment of $c_s^2 = 1$ followed by a phase transition or vice versa, respectively. 
These EOSs bound the range of possible pressure differences to be 
\begin{align}
    \Delta p_{\min} & = \frac{n_L}{2} 
    \left(
    \frac{\mu_H^2}{\mu_L}-\mu_L 
    \right),  \\
    \Delta p_{\max} & = \frac{n_H}{2} 
    \left(
    \mu_H -\frac{\mu_L^2}{\mu_H}
    \right).
\end{align}

Given values for $\beta_L$ and $\beta_H$, the pressure difference between these two points is fixed. 
Therefore the simple check (for detailed derivation see Refs.~\cite{Komoltsev:2021jzg,Gorda:2022jvk})
\begin{align}
 \Delta p = p_H - p_L & \in [\Delta p_{\min}, \Delta p_{\max}] 
\end{align}
determines whether the two points can be connected by a causal and stable EOS. 
This condition does not make use of any specific form of the EOS between the low- and high-density points but rather implies the existence of at least one EOS connecting the two limits. 
The condition on the pressure difference can be rewritten as
\begin{align}
0 \le I_\pQCD \equiv \frac{\Delta p - \Delta p_{\min}}{\Delta p_{\max} - \Delta p_{\min}} \le 1\,,
\label{eq:tension}
\end{align}
where we introduce a new quantity $I_\pQCD$, the pQCD tension index. 
If the index $I_\pQCD \notin [0,1]$ the EOS is excluded because there is no possible causal and stable interpolation. 
If the index obtains the values $I_\pQCD=0  \,(I_\pQCD=1)$, then the EOS is forced to follow $n_\mathrm{pmin} \,(n_\mathrm{pmax})$, respectively. 
In particular, if $I_\pQCD = 1$, the EOS must contain a phase transition of size
\begin{equation}
\Delta n = n_H \frac{\mu_L}{\mu_H} - \mu_L\,.
\end{equation}
For typical values of $n_L=n_{\term} = n_\tov$, we obtain that the strength of phase transition reaches $\Delta n \sim 20 n_\sat$, followed by a  $c^2_s$ = 1 segment of $\sim 10 n_\sat$. 
Note that if $I_\pQCD$ is not exactly $I_\pQCD=0$ or $I_\pQCD=1$, then both extreme EOSs of Eqs.~\eqref{npmin} and \eqref{npmax} are excluded. 
For $0 < I_\pQCD < 1$, the EOS is bounded in the $n$--$\mu$--$p$ -space by a non-trivial envelope whose projection in the plane of energy-density $\varepsilon$ and pressure $p$ is shown in Fig.~\ref{fig:1} (for an explicit expression, see Ref.~\cite{Komoltsev:2021jzg}).

Following Ref.~\cite{Gorda:2022jvk}, we adopt the scale-averaging interpretation of the pQCD uncertainties for $\beta_H$. 
We use a log-uniform distribution for the dimensionless renormalization scale $X = 3 \bar\Lambda/2\mu_\pQCD$ in the range [1/2, 2], which can be written as $w(\log X) =  {\mathbf 1}_{[ \log(1/2),\,\log(2)]}(\log X)$. 
Here $\bar\Lambda$ is the renormalization scale in the modified minimal subtraction scheme~\cite{Kurkela:2009gj}.
The QCD likelihood function then reads
\begin{align}
\label{eq:QCD-likelihood}
P( {\rm QCD| EOS} ) = \int &d(\log X) w(\log X) \times \nonumber \\ &\mathbb{1}_{[0,1]}\bigl(I_\pQCD(X,\rm EOS) \bigr)
\end{align}
where $\mathbb{1}_{[0,1]}(I_\pQCD(X,\rm EOS))$ is an indicator function discarding EOSs with $I_\pQCD \notin [0,1]$ for a fixed value of $X$.
In general, the index $I_\pQCD$ depends on all six quantities $(n_L, p_L, \mu_L)$ and $(n_H, p_H, \mu_H)$, which for fixed values of $\mu_H=\mu_{\rm QCD}$ reduces to $I_\pQCD(X,\mathrm{EOS})= I_\pQCD(X,n_L, p_L,\mu_L)$.

We note that the convergence of the pQCD series depends on the unphysical renormalization scale. 
In particular, the convergence is slowest for the smaller values of $X\approx 1/2$. 
This slow convergence renders the $X=1/2$ results, which differ quantitatively from the higher values of $X$, less trustworthy; see Ref.~\cite{Gorda:2023usm} for a detailed analysis of the convergence of the pQCD results.

In the following sections we apply the QCD likelihood function at different termination densities, meaning that $(n_L, p_L, \mu_L)$ corresponds to $(n_{\term}, p_{\term}, \mu_{\term})$. 
For the high density input, $\beta_H=\beta_\pQCD$, we use the partial next-to-next-to-next-to-leading order pQCD computation of Ref.~\cite{Gorda:2021znl}. 

Note that in terms of the sound speed, the QCD input constrains the density $n_\term$, which is given by a (logarithmic) integral of $c_s^{-2}(\mu)$ and the pressure $p_\term$, which is the integral of this integral $n(\mu)$.
Hence for a given NS EOS, the input is more sensitive to individual features in the speed of sound at lower densities and is less sensitive to those near $n_\term$.

\section{Computational Setup}
\label{sec:computational_setup}

In order to study the dependence of our results on the differences in the analyses between Refs.~\cite{Gorda:2022jvk,Somasundaram:2022ztm}, we use two different models to generate our EOS sets: the parametric sound-speed model of Ref.~\cite{Somasundaram:2022ztm} and the Gaussian-process (GP) model of the speed of sound of Ref.~\cite{Gorda:2022jvk}. 
Note that similar models have also been employed by Refs.~\cite{Somasundaram:2021clp,Tews:2018kmu,Landry:2018prl,Mroczek:2023zxo}.

\subsection{Prior Set 1: The sound-speed model}

To generate the CSM set, we follow Ref.~\cite{Somasundaram:2022ztm}. 
At the lowest densities, we fix the EOS to the crust model of Ref.~\cite{Douchin:2001sv} and do not explore uncertainties in the crust modeling. 
The crust EOS is then matched with the EOS of the core via a cubic spline in the $c^2_s$ versus $n$ plane.
For the core EOS, below a certain breakdown density $n_b$, we assume that the EOS can be accounted for by nucleonic degrees of freedom, which in this work is described by the meta-model of Refs.~\cite{Margueron2018a,Margueron2018b}. 
The meta-model is a density-functional approach, similar to the Skyrme approach, that allows one to incorporate nuclear-physics knowledge directly encoded in terms of the Nuclear Empirical Parameters (NEPs). 
We vary the NEPs in a rather broad range in order to explore a wide range of nuclear density functionals that describe matter up to $n_b$. 
The parameter $n_b$ is varied uniformly in the range $1-2 n_\sat$. 
Furthermore, we discard all models that do not satisfy constraints provided by the chiral EFT calculations of Ref.~\cite{Lynn:2015jua} up to nuclear saturation density.

Above $n_b$, we use the CSM and create a non-uniform grid in density between $n_b$ and $25 n_\textrm{sat}$ with $9$ grid points, where the 9th point is fixed at $25 n_\textrm{sat}$ and the other 8 points are drawn from a uniform probability distribution in density.
At each grid point, the squared sound-speed is uniformly sampled between $0$ and $1$ in units of the speed of light. 
Then, the sound speeds at each grid point are connected by linear segments, generating a sound-speed profile $c^2_s(n)$ for the EOS. 
The density-dependent speed of sound $c^2_s(n)$ can then be integrated to give the EOS, i.e., the pressure $p(n)$, the energy density $\varepsilon(n)$, and the baryon chemical potential $\mu(n)$, see Ref.~\cite{Somasundaram:2021clp} for more details. 
The CSM set used here consists of $\approx16,000$ EOSs generated this way.

\subsection{Prior Set 2: Gaussian-Process model}

In the GP model, for densities below $n = 0.57n_\sat$ we use the crust model by Baym, Pethick, and Sutherland~\cite{Baym:1971pw}. 
Above the crust-core transition density, we perform a GP regression in an auxiliary variable $\phi(n)~\equiv~-\ln(1/c_s^2(n) - 1)$, choosing the prior on this variable from the multivariate Gaussian distribution ,
\begin{equation}
\phi(n) \sim \mathcal{N}\bigl(-\ln (1/\bar{c}_s^2 - 1), K(n, n')\bigr)\,,
\end{equation}
with a Gaussian kernel $K(n, n') = \eta \exp \bigl( - (n - n')^2 / 2 \ell ^2 \bigr)$ \cite{Gorda:2022jvk}. 
In our hierarchical model the hyperparameters of the GP $\eta$, $\ell$, and $\bar{c}_s^2$ are themselves drawn from normal distributions,
\begin{equation}
\begin{split}
\eta& \sim \mathcal{N}(1.25, 0.2^2)\,, \\
\ell& \sim \mathcal{N}(1.0 n_\sat, (0.25 n_\sat)^2)\,, \\
\bar{c}_s^2& \sim \mathcal{N}(0.5, 0.25^2)\,.
\end{split}
\end{equation}
For $n \leq 1.1 n_\sat$, this prior is then conditioned on the chiral EFT results from Ref.~\cite{Hebeler:2013nza}, with the average and difference between the ``soft'' and ``stiff'' EOSs in that work taken to be the mean and 90\% credible interval for the conditioning~\cite{Gorda:2022jvk}.
The final EOSs are then once again obtained by integrating $c_s^2(n)$.
The GP used in this work consists of 120,000 EOSs generated in this manner.

\begin{figure*}[h!]
\includegraphics[scale = 0.55]{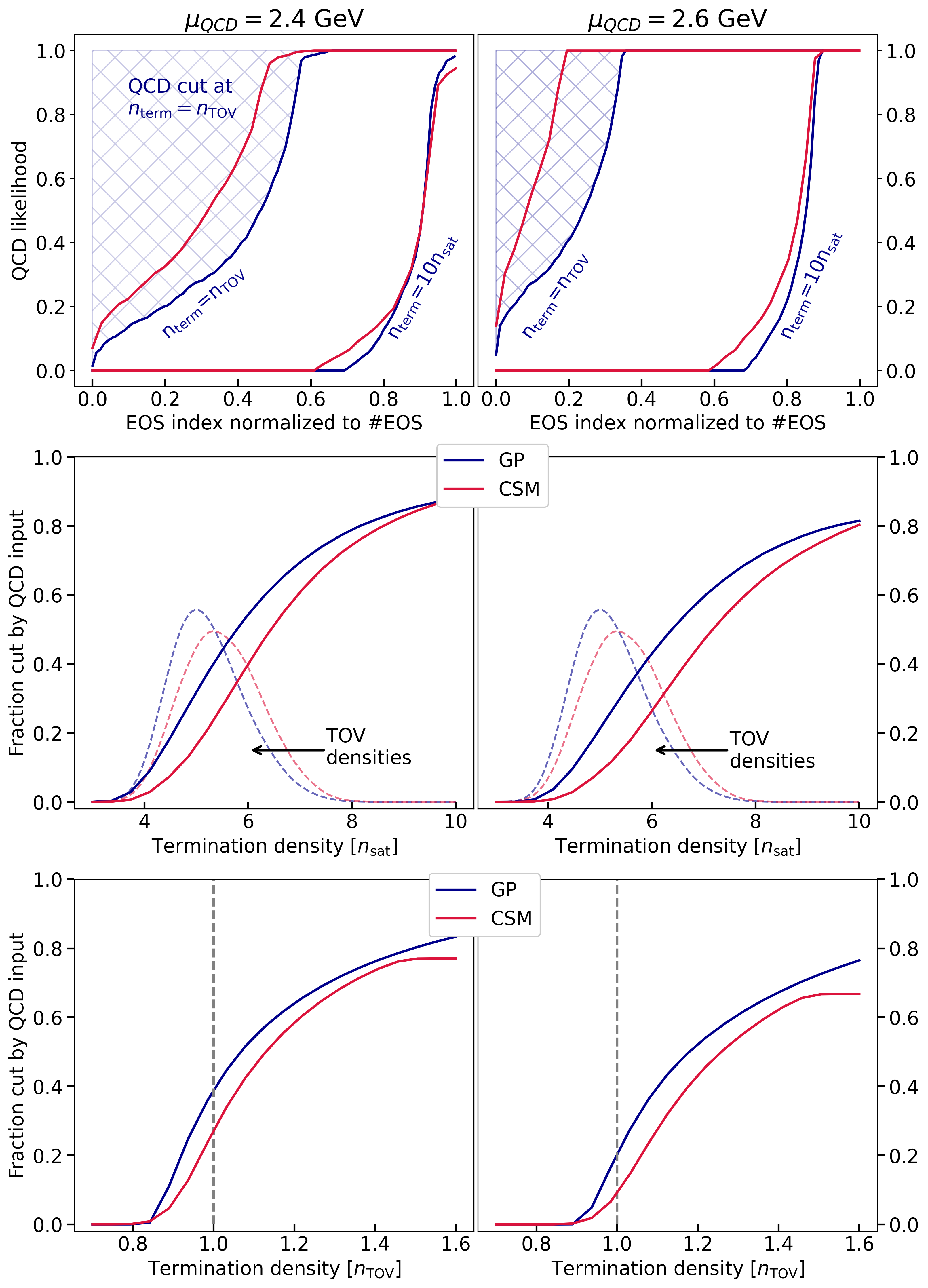}
\caption{(\textbf{Top}) The QCD likelihood function for $\mu_{\rm{QCD}}=2.4$~GeV (left panels) and $\mu_{\rm{QCD}}=2.6$~GeV (right panels) for EOS posteriors conditioned on astrophysical data.
The EOSs are ordered according to their QCD likelihood.
The blue lines correspond to the GP, whereas the red lines correspond to the CSM.  
The likelihood function is displayed for two different termination densities, $n_{\term}=n_\tov$ and ${n_{\term}=10n_\sat}$, labeled in the figure. 
(\textbf{Middle}) Fraction of the astrophysical evidence (see Eq.~\ref{eq:fraction_cut}) that is cut by the QCD input as a function of the termination density. 
The dashed blue distribution shows the posterior for the TOV densities for the GP whereas the dashed red lines show the same for the CSM. 
(\textbf{Bottom}) Fraction of the astrophysical evidence (see Eq.~\ref{eq:fraction_cut}) that is cut by the QCD input as a function of the termination density in units of the TOV density.
}
\label{fig:likelihood-panels}
\end{figure*}

\subsection{Astrophysical input}

The astrophysical inputs that we consider in our analysis are the following:
\begin{enumerate}
\item Radio-astronomy constraints: 
We consider the mass-measurements of PSR J0348+0432 with $M = 2.01 \pm 0.04 M_\odot$~\cite{Antoniadis:2013pzd} and PSR J1624$-$2230 with $M = 1.928 \pm 0.017 M_\odot$~\cite{Fonseca:2016tux}. 
We approximate these uncertainties as normal distributions, which is valid to good accuracy.

\item Gravitational-wave data from binary NS mergers: We consider the two-dimensional PDF for the tidal deformability and mass ratio at fixed chirp mass extracted from GW170817 by the LIGO-Virgo collaboration~\cite{LIGOScientific:2018hze}, using the PhenomPNRT waveform model with low-spin priors (see Figure 12 of \cite{LIGOScientific:2018hze}).

\item NICER observations: We consider the simultaneous mass and radius measurement of PSR J0740$+$6620 using the NICER + XMM-Newton data. 
We have employed here the PDF from Ref.~\cite{Miller:2021qha} (see their Fig.~1, right panel).
\end{enumerate}

After conditioning our priors with the astrophysical input, we obtain a posterior sample of 4,000 EOSs for CSM and 46,000 EOSs for the GP model.
These inputs are identical to those used in Ref.~\cite{Gorda:2022jvk} (referred to as ``Pulsars + $\tilde\Lambda$'') with one exception. 
We choose not to use the additional ``Occam'' factor related to the prior mass distribution of NSs, ${(m_{\tov} - m_{\min})^{-4}}$ (one power per mass integral), which penalizes high-mass EOSs that are completely consistent with astrophysical data; see Ref.~\cite{Miller:2019nzo,Landry:2020vaw} for a discussion of this point.
These inputs are the same as used in Ref.~\cite{Somasundaram:2022ztm} with the exception of the NICER measurement of PSR~J0030+0451. 
We have not included the latter in the present analysis as we found its impact on estimating the pQCD constraining power to be small.

Ref.~\cite{Gorda:2022jvk} analyzed the astrophysical as well as the QCD inputs using a Bayesian approach whereas Ref.~\cite{Somasundaram:2022ztm} performed a hard cut analysis. 
We have checked that, as far as the treatment of the astrophysical data is concerned, this difference does not alter the conclusions of our paper regarding the impact of the QCD input. 
The role played by the different treatments of the QCD input is discussed in the next section.  

\section{Results}
\label{sec:results}

Next, we focus on the impact of the differences in methodology between Refs.~\cite{Gorda:2022jvk} and~~\cite{Somasundaram:2022ztm}.
The three major differences between the setups in these works can be summarized as follows:
\begin{enumerate}
\item[D1:] Ref.~\cite{Gorda:2022jvk} used a fully Bayesian approach and the predictions are analyzed in terms of posterior probability distributions whereas Ref.~\cite{Somasundaram:2022ztm} performed a hard cut analysis and focused on the envelope associated with all EOS models for all studied quantities.

\item[D2:] The functional basis of the EOS modeling in the two works was different: Ref.~\cite{Gorda:2022jvk} employed the non-parametric GP model (Prior set 2), while  Ref.~\cite{Somasundaram:2022ztm} employed the parametric CSM (Prior set 1). 

\item[D3:] The two works made different choices for $n_\mathrm{term}$: 
Ref.~\cite{Gorda:2022jvk} chose $n_\mathrm{term} = 10n_\sat$ whereas Ref.~\cite{Somasundaram:2022ztm} chose $n_\mathrm{term} = n_\tov \approx 5-8n_\sat$, which is different for each EOS. 
Hence, the works inferred the EOS in different density ranges.
\end{enumerate}
To investigate point D3, we will discuss results for varying $n_\term$ expressed either in terms of multiples of the saturation density $n_\sat$ or in terms of the EOS-dependent central density of the maximum-mass neutron star, $n_\tov$. 

\subsection{Impact of the QCD input on the EOS}

\begin{figure}[b]
\includegraphics[width = \columnwidth]{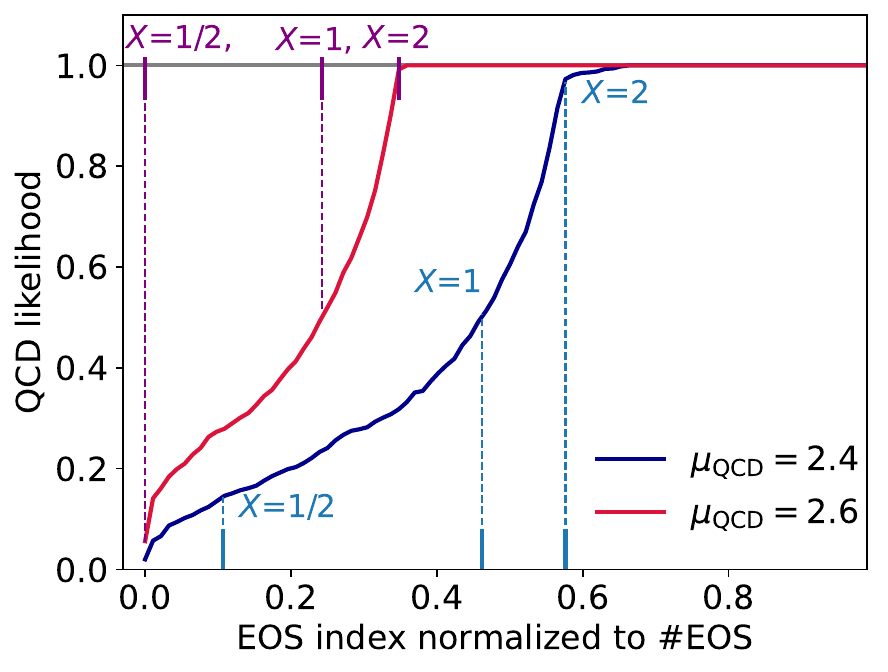}
\caption{ The QCD likelihood function at $n_{\term}=n_{\tov}$ for the GP EOS posteriors conditioned on astrophysical data. The EOSs are ordered according to their QCD likelihood. 
The blue line corresponds to $\mu_{\pQCD}=2.4$ GeV and a scale-averaged prescription for $X$ in the range [1/2,2], whereas the red line corresponds to $\mu_{\pQCD}=2.6$ GeV. 
The colored ticks and dotted lines correspond to the likelihood function for fixed values of $X$ ($X$=1/2, 1 and 2), taking the form of a step function. 
}
\label{fig:diffx}
\end{figure}

We show the QCD likelihood of Eq.~\eqref{eq:QCD-likelihood} for the two EOS models, GP and CSM, in the upper panels of Fig.~\ref{fig:likelihood-panels}, where both models terminate at $n_\term = n_{\tov}$. 
In particular, we show the sorted QCD likelihood for a large sample of EOSs drawn from the posterior distributions of the CSM and GP models after they have been conditioned on the astrophysical data. 
EOSs for which the likelihood is $1$ are accepted for all values of the pQCD renormalization scale $X$ whereas EOSs that have a likelihood of $0$ cannot be connected to pQCD with a causal and stable interpolation for any value of $X \in [1/2,2]$.
EOSs with likelihoods between the two limits can be connected to pQCD for some but not all values of $X$, see Fig.~\ref{fig:diffx} for more details. 
Figure~\ref{fig:likelihood-panels} clearly shows that the choice of EOS model has a weak impact on the QCD likelihood, illustrating that the difference D2 plays a minor role in the following analysis.

We observe that for $n_\term = n_\tov$ and $\mu_H = 2.6$~GeV ($\mu_H = 2.4$~GeV) roughly 20\% ($\sim 40 \%$) of the EOSs cannot be connected to pQCD for all values of $X$ and are therefore affected by the QCD input.
For a very small fraction of EOS, the QCD likelihood drops below $20\%$ but no EOS has a vanishing QCD likelihood for this choice of $n_\term$. 
In contrast, the results are very different for $n_\term = 10n_\sat$, which was chosen in Ref.~\cite{Gorda:2022jvk}.
Now, roughly $50-60\%$ of EOS have a vanishing QCD likelihood.
This shows the strong impact of the choice of $n_\term$ on the pQCD constraining power and explains a large part of the difference in the conclusions of Refs.~\cite{Gorda:2022jvk, Somasundaram:2022ztm}.
The strong differences between the two cases arise from the behavior the EOS must adopt beyond $n_\tov$ to remain consistent with the QCD input at $10 n_\sat$. 
This is discussed further in Section~\ref{sec:IVC}.

We quantify the impact of the QCD likelihood function as a function of $n_\term$ by computing the fraction of the marginalized likelihood (evidence) excluded by the QCD input 
\begin{align}
    1-\frac{\sum_i w^{\rm astro }_i \cdot w^{\rm QCD}_i}{\sum_i w^{\rm astro }_i}\,,
    \label{eq:fraction_cut}
\end{align}
where the index $i$ denotes the EOS, and the summation is performed over all EOSs in the ensemble. 
The symbols $w^{\rm astro}$ and $w^{\rm QCD}$ correspond to the likelihood arising from the astrophysical and QCD inputs, respectively. 
In Ref.~\cite{Pang:2023dqj} this quantity was called the average consistency (see also Eq.~(25) of Ref.~\cite{Gorda:2022jvk}), and corresponds to one minus the average QCD likelihood in the posterior ensemble conditioned with only the astrophysical input, which is indicated by the hatched area.

\begin{figure*}[t]
\includegraphics[scale = 0.5]{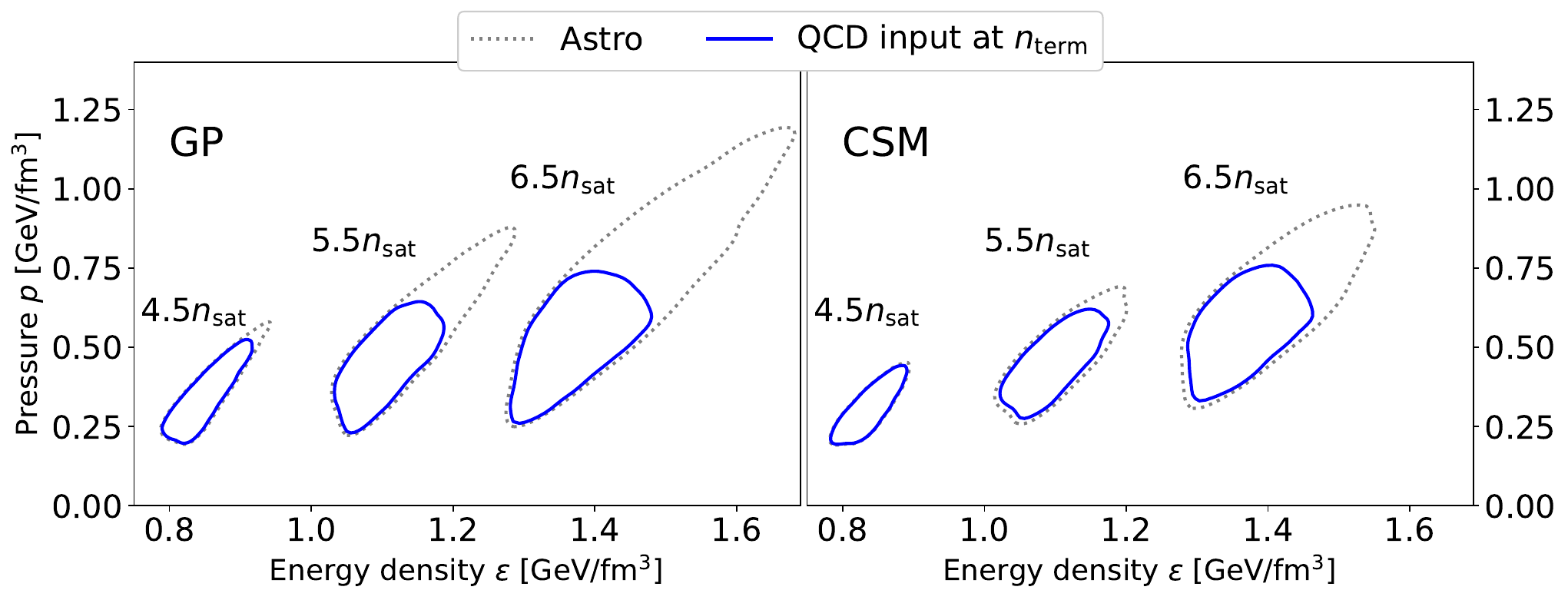}
\caption{Posterior probability density of the energy density vs.~pressure with the QCD input enforced at three different fixed termination densities. 
The average TOV density for the GP posterior after enforcing astrophysical information is approximately 5.5$n_\sat$, while the maximum TOV density reaches approximately 8$n_\sat$ for the softest EOS. 
The areas correspond to 68\% credible regions when considering only  astrophysical information (dotted gray) and when also including the QCD input at $\mathrm{n_{term}}$ (solid blue). 
The QCD input is imposed at 2.6 GeV with a scale-averaging prescription. }
\label{fig:e-p_panel2}
\end{figure*}

This quantity is shown as the function of the termination density for the two EOS models in multiples of $n_\sat$ in the middle panels and relative to $n_{\tov}$ in the bottom panels of Fig.~\ref{fig:likelihood-panels}.
As the models are extended to higher $n_\term$, the fraction of the astrophysical evidence cut by QCD increases rapidly; if the models are extended to $n = 1.2 n_{\tov}$ this fraction increases to 0.4 (0.5 for $\mu_\mathrm{QCD} = 2.4$ GeV). 
At a fixed density of $n_\term = 6.5n_\sat$, roughly corresponding to the approximate 68\%-credible upper limit for the $n_{\tov}$, the average consistency is 0.4. 
However, we stress that the stiffest EOSs, i.e., those with the lowest QCD likelihoods, are in the lower percentile of the $n_{\tov}$ range, and hence, for these EOS $n_{\tov}<6.5n_\sat$.
These EOS are more strongly disfavored for larger $n_\term$, when the EOS branch beyond $n_{\tov}$ is probed.
For higher termination densities, the average consistency levels out and takes values of $\sim 0.8$ at $n_{\term} = 10 n_\sat$. 

We, hence, find that the constraining power of the QCD input depends largely on the choice of the termination density $n_\term$ (D3).
In fact, we find that a value of $n_\term \approx 0.8 n_\tov$ ($n_\term \approx 4 n_\sat$ when applied at a fixed density) constitutes a critical value, in the sense that it is at this point that the QCD input begins to show any effect at all beyond the astrophysical input.
We note that the fact that this critical value of $n_\term$ is so close to $n_\tov$ is a coincidence, in the sense that this critical value of $n_\term$ is set by a combination of many factors, including the value of $n_\pQCD$ at which pQCD is well-converged and the constraining power of the astrophysical input (the latter applying only below $n_\tov$). 
In particular, if $n_\pQCD$ were significantly lower, then it is possible that this critical value of $n_\term$ would also be lower.

In order to investigate the impact of point D1 above, we show in Fig.~\ref{fig:diffx} the detailed impact of $X$ on the QCD likelihood.
For different values of $X$, we indicate where the likelihood function becomes 1, i.e., all EOS to the right of these lines are accepted by the QCD input at $n_\term = n_\tov$, while those on the left side are ruled out.
Considering $\mu_\pQCD=2.6$~GeV as in Refs.~\cite{Gorda:2022jvk,Somasundaram:2022ztm}, we find that nearly all EOS are accepted by $X=1/2$, recovering the results of Ref.~\cite{Somasundaram:2022ztm}. 
In contrast, roughly 25\% of EOSs are rejected for $X=1$ and 60\% for $X=2$. 
If $\mu_\pQCD$ is decreased to $\mu_\pQCD=2.4$~GeV, about 10\% of EOS are rejected already at $X=1/2$. 
A Bayesian treatment of the scale parameter $X$ as performed in Ref.~\cite{Gorda:2022jvk} averages the impact of $X=1/2$ leading to the findings of Ref.~\cite{Gorda:2022jvk}. 
Hence we conclude that different statistical treatment of including pQCD calculations (D1) plays a secondary role in the differing conclusions of the two studies.

\subsection{\texorpdfstring{Impact of the QCD input on the $\varepsilon$--$p$ values}{Impact of the QCD input on the e,p values}}

\begin{figure*}
\includegraphics[scale = 0.5]{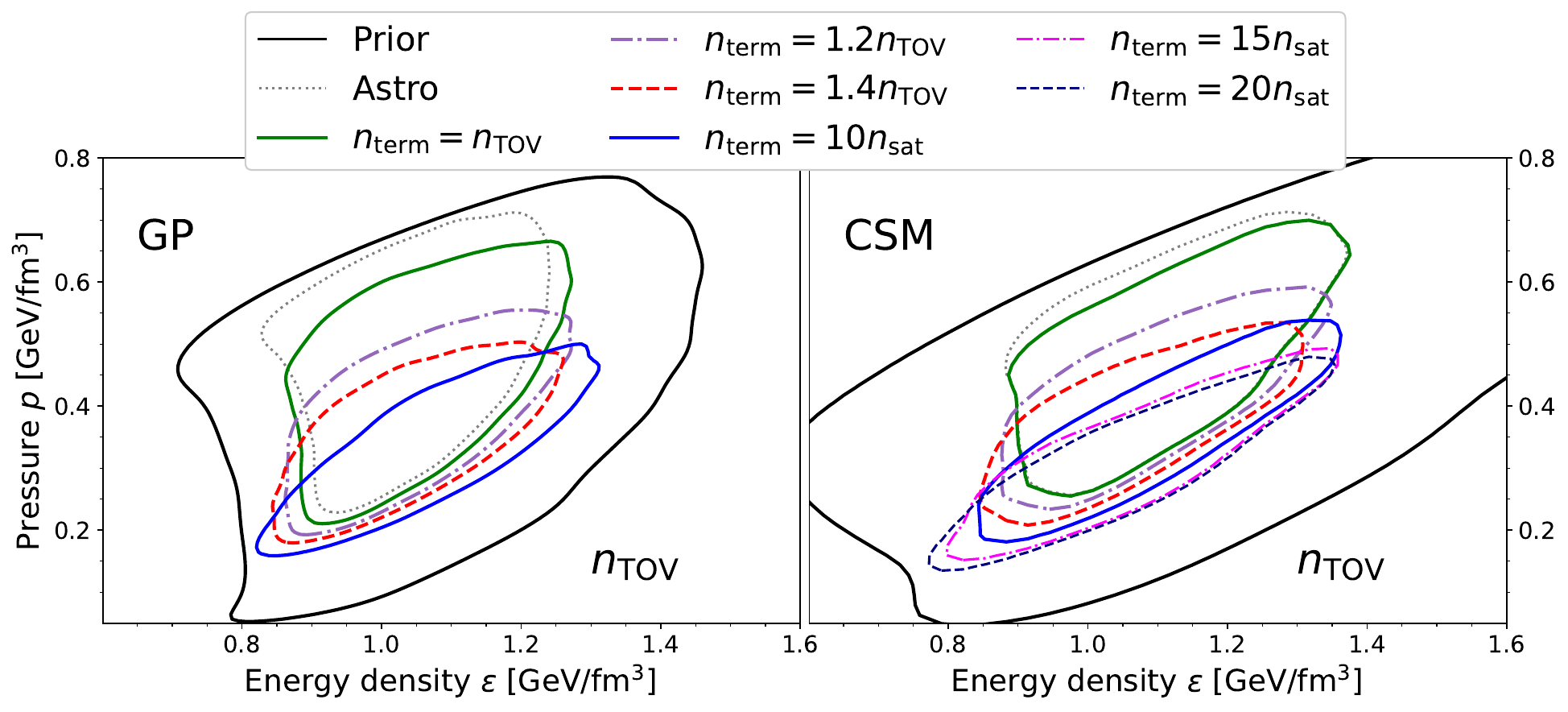}
\caption{Posterior probability density of the energy density vs. pressure at $n_\tov$ for different termination densities. 
Each contour corresponds to 68\% credible regions for the prior (black solid), considering only astrophysical input (dotted gray) and when also including QCD input (different colors corresponds to different termination densities). 
The left plot shows results using the GP approach, while the right one shows results for the CSM approach. The QCD input is imposed at 2.6 GeV with a scale-averaging prescription.}
\label{fig:e-p_panel1}
\end{figure*}

We now analyze the effect of the QCD input on the allowed values of energy density and pressure. 
Figure~\ref{fig:e-p_panel2} shows posterior credible regions of energy density $\varepsilon$ and pressure $p$ at various fixed densities in units of saturation density, incorporating either the astrophysical input alone or additionally also the QCD input. 
The EOS models are terminated at different densities $n_\term \in \{4.5, 5.5, 6.5\}n_\sat$, which correspond roughly to the range of $n_\tov$ shown above in Fig.~\ref{fig:likelihood-panels}. 
We note that taking the termination density to be the density we are studying represents the maximally conservative approach to applying QCD constraint at a fixed density, as the QCD input is completely independent of the prior beyond $n_\term$. 
As discussed in Ref.~\cite{Gorda:2022jvk}, the main effect of the QCD input is to disfavor the stiffest EOSs, visible in the figure. 
The effect of the input depends on the density; consistent with Fig.~\ref{fig:likelihood-panels}, at densities $\sim 6.5 n_\sat$ (corresponding to the 68\%-credible upper limit on $n_\tov$) there is a significant reduction of allowed $\varepsilon$--$p$ values, particularly at the highest $p$ and $\epsilon$.
At lower densities the constraint is weaker as expected; at 5.5$n_\sat$, corresponding to the mean of $n_\tov$, the reduction is still significant, while taking $n_\term  =4.5n_\sat$ only shifts the posterior slightly at the highest pressures. 
This is in line with the prediction of Ref.~\cite{Komoltsev:2021jzg} stating that QCD input ceases to give any constraint at $n < 2.2-2.5 n_\sat$.  
From Fig.~\ref{fig:e-p_panel2}, we can also clearly see a difference between the two NS EOS models, namely that the GP model gives more weight to EOSs with larger values of $p$ and $\epsilon$ than the CSM, visible in the posterior for the astrophysical input. 
This explains the slightly larger effect of the QCD input within the GP for the same $n_\term$ (point D2 above).
However, after the QCD input is applied, the two models give quite similar posteriors.

We now move from studying the EOS at fixed densities to varying the termination density in terms of $n_{\tov}$. 
In Fig.~\ref{fig:e-p_panel1}, we show the posterior distribution of energy density $\varepsilon(n_{\tov})$ and pressure $p(n_{\tov})$ for $n_\term \in \{1, 1.2, 1.4\} n_{\tov}$.   
The figure shows the 68\%-credible regions for the prior distribution before introducing the astrophysical constraints as well as the regions after the EOSs have been conditioned on the astrophysical inputs but before the QCD input has been imposed. 
For both EOS models, the upper (left) edge of the credible region contour is nearly unchanged by the astrophysical input. 
This demonstrates that there is a region in the $\varepsilon(n_{\tov})$--$p(n_{\tov})$---plane that is only marginally constrained by the present astrophysical input. 

Similar to the $\varepsilon$--$p$---distribution at fixed density, the effect of the QCD input is to soften the EOS, disfavoring the highest pressures. 
We observe that for $n_{\term} \in \{1.2, 1.4\} n_{\tov}$ the QCD input sets the upper left edge of the posterior distribution. 
This is also the case for $n_\term = n_\tov$ in the GP model, while for the CSM at $n_{\term} = n_{\tov}$ the impact is marginal. 
Consistent with Fig~\ref{fig:likelihood-panels}, the quantitative impact is smallest for $n_{\term} = n_{\tov}$ but the impact grows rapidly when $n_{\term}$ is increased and is qualitative similar to setting $n_{\term}$ to fixed density $10 n_\sat$, shown in Fig.~\ref{fig:e-p_panel1}. 
From both of these two figures, we observe that the choice of EOS prior (either GP or CSM) has only a minor impact on the effect of the QCD constraint. 
Finally, we also show contours corresponding to a fixed $n_{\term} \in \{10, 15, 20\} n_{\sat}$, where we show only CSM contours for the latter two values.
We note that applying the QCD input for $n_\term \in \{10, 15, 20\} n_\sat$ yields very consistent credible regions.

\begin{figure*}
\centering
\includegraphics[scale = 0.46]{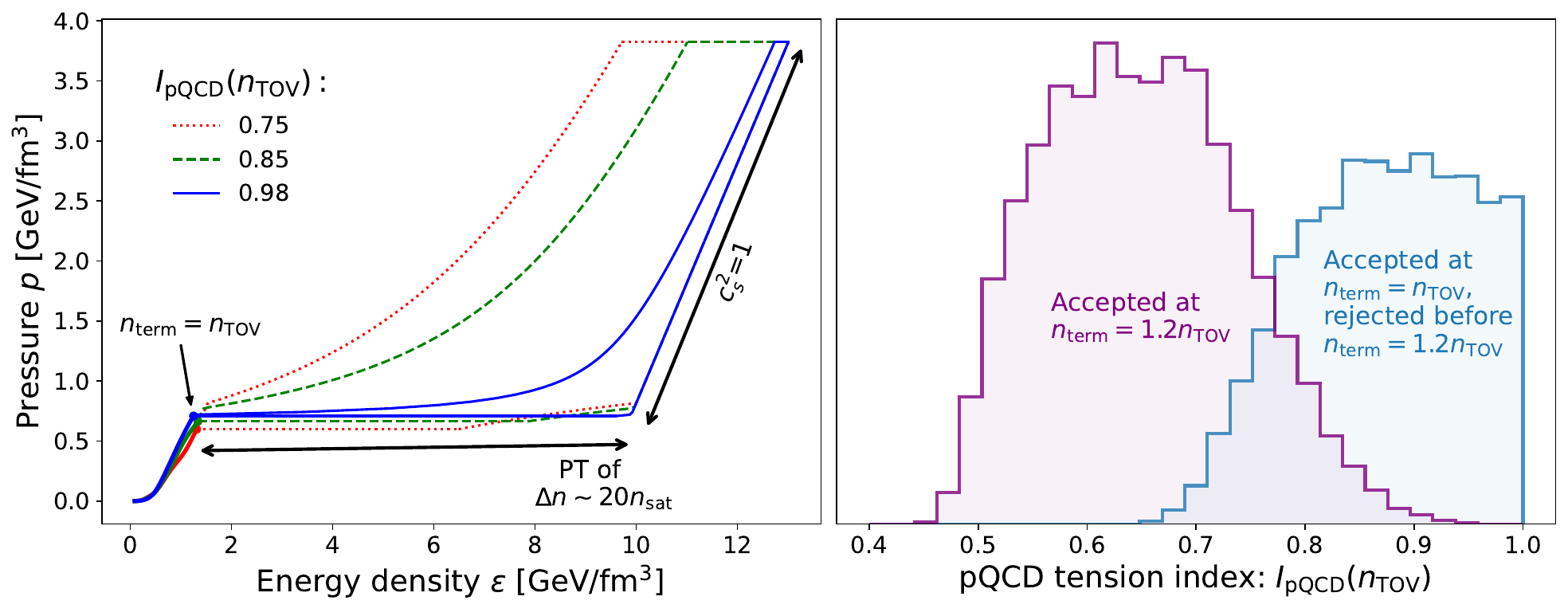}
\caption{ (\textbf{Left}) The QCD constraint shown in Fig.~\ref{fig:1} for 3 different EOS with representative values of the pQCD tension index (see Eq.~\ref{eq:tension}).
Each EOS is terminated at $n_{\term}=n_{\tov}$ and accepted by QCD input for a fixed $X$=1 and $\mu_{\rm QCD}=2.6$ GeV.
The arrows illustrate the phase transition and the subsequent $c_s^2=1$ segment, which an EOS is forced to have when the tension index is close to 1. 
(\textbf{Right}) The distribution of the pQCD tension index for EOSs that are accepted at $1.2 n_{\tov}$ and, therefore, also at $n_{\tov}$ (purple) as well as for EOS that are accepted at $n_{\tov}$, but rejected somewhere between $n_{\tov}$ and $1.2 n_{\tov}$ (blue).
These distributions are normalized to the number of EOSs in each distribution.
Most of the EOSs from the latter ensemble have a large pQCD tension index. 
}
\label{fig:tesion}
\end{figure*}

\begin{figure}
\centering
\includegraphics[scale = 0.53]{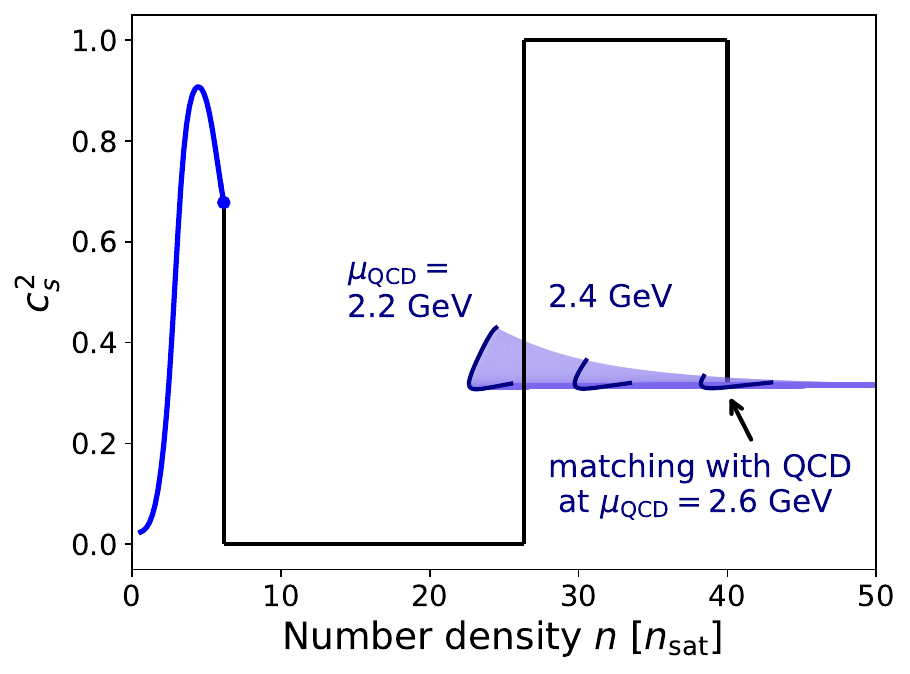}
\caption{The behavior an EOS with tension index $I_\pQCD = 1$ at $n_{\rm TOV}$ (blue line) has to follow above the termination density $n > n_\term$ to match with pQCD for $X = 1$ (black line).
The purple band corresponds to pQCD calculations of the speed of sound starting from $\mu_{\rm QCD}=2.2$ GeV, with $X$ varying in the range [1/2,2]. 
}
\label{fig:extension_cartoon}
\end{figure}

\begin{figure*}
\centering
\includegraphics[scale = 0.53]{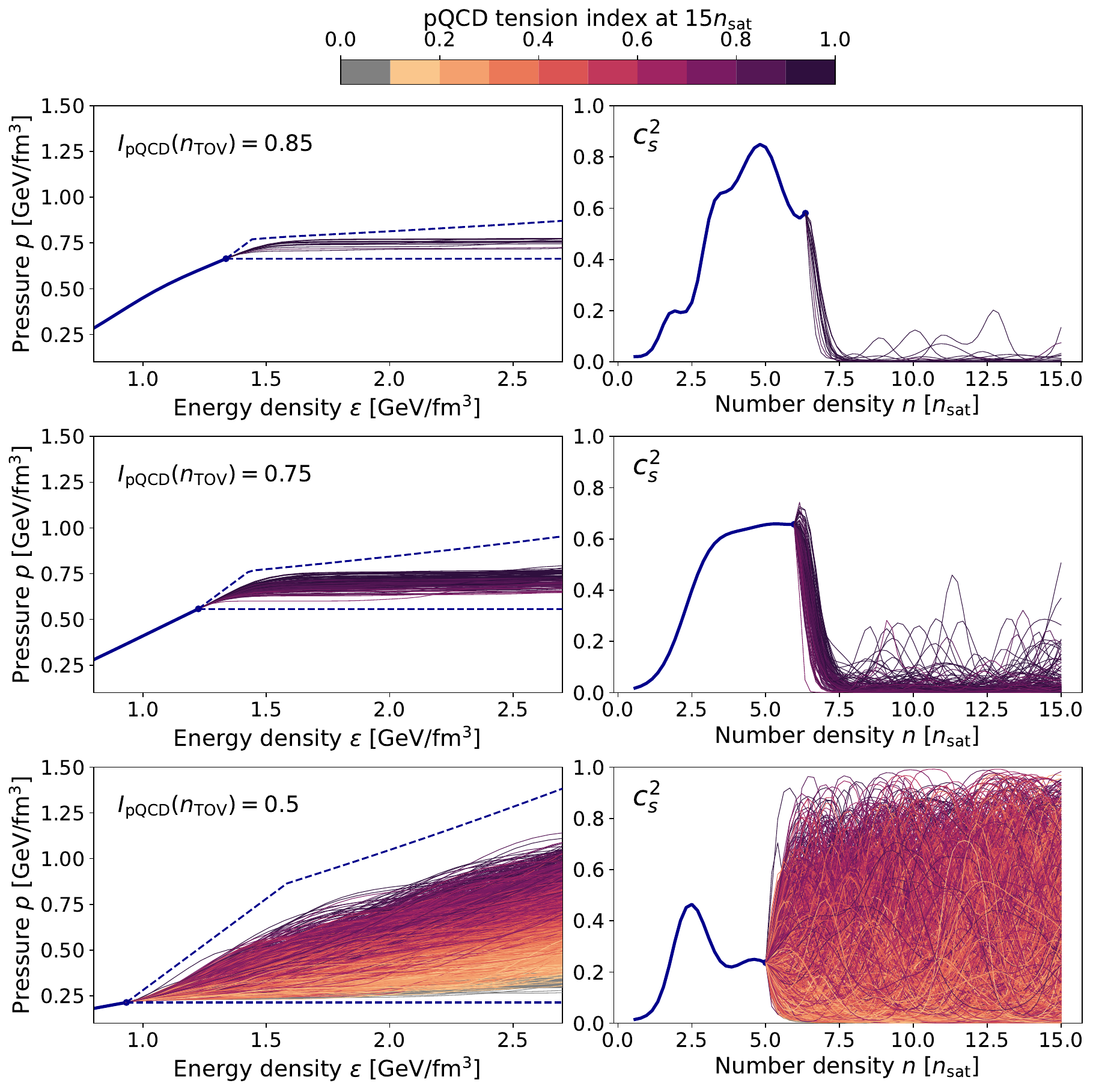}
\caption{Ensembles of possible extensions for three representative EOSs that are allowed at $n_{\term} = n_\tov$. 
The EOSs correspond to different values for the pQCD tension index $I_{\rm pQCD}(n_\tov)$. 
The extensions are conditioned to fulfill the QCD constraint at 15$n_s$.
In this figure, the coloring of each EOS is given by the value of $I_\pQCD(15 n_\sat)$.}
\label{fig:extension_ensemble}
\end{figure*}

\begin{figure}
\centering
\includegraphics[scale = 0.48]{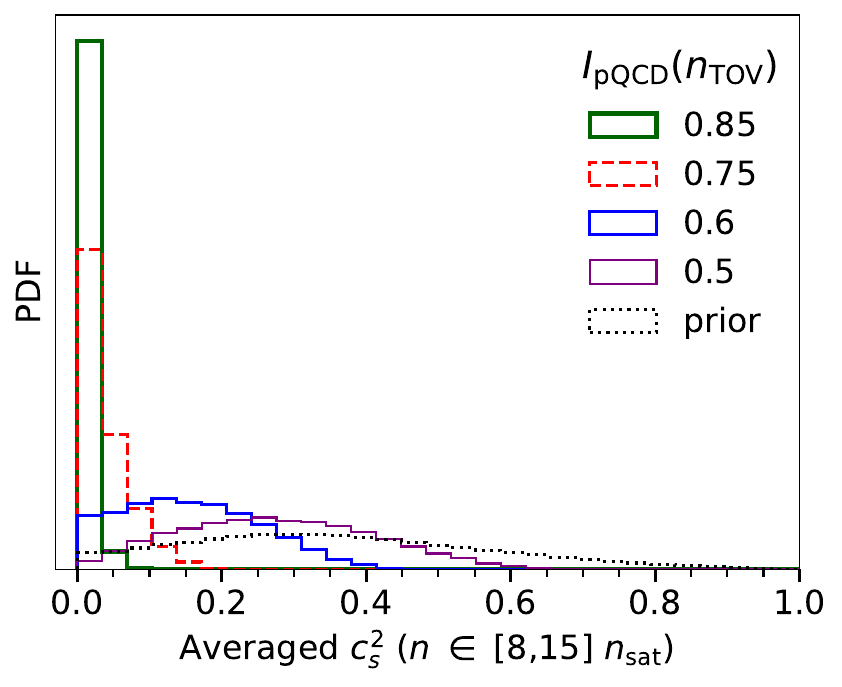}
\caption{The distributions of the averaged sound speed in the density interval $[8,15]n_{\rm sat}$ for the possible EOS extensions of 100 different EOSs up to $n_\term = n_\tov$. 
For each EOS, we generate 1000 extensions.
This distribution is shown for fixed values of the pQCD tension index $I_\pQCD(n_\tov) = 0.85, 0.75, 0.6, 0.5$.}
\label{fig:change_of_index}
\end{figure}

\subsection{\texorpdfstring{Implications for the EOS beyond $n_{\term}$}{Implications for the EOS beyond n\_term}}
\label{sec:IVC}

So far, we have discussed the impact the QCD input has on the EOS for $n < n_\term$. 
In this subsection, we discuss how the QCD constraint affects the behavior of the EOS at densities $n > n_\term$.
As shown in Fig.~\ref{fig:1}, the QCD input affects the available parameter space for the EOS in the $p$--$\varepsilon$ plane beyond $n_{\term}$, $n_{\term} < n < n_H$.
Some EOS have a large available parameter space while others are tightly constrained. 
One possible measure to gauge this freedom is the pQCD tension index $I_\pQCD$ defined before. 
The left panel of Fig.~\ref{fig:tesion} shows the allowed space in the $\varepsilon$--$p$ plane beyond $n_\term$ for three EOSs with $I_\pQCD(n_\tov) = 0.75, 0.85$, and $0.98$. 
The EOS with the highest $I_\pQCD$ must undergo a rapid phase-transition-like softening with $\Delta n \sim 20 n_\sat$, followed by a segment of $c_s^2 \approx 1$ to remain consistent with pQCD calculations. 
For $I_\pQCD=1$, this discontinuous behavior would be be completely determined -- see Fig.~\ref{fig:extension_cartoon}. 
Note that this phase transition is not responsible for the onset of the unstable NS branch, which instead is determined by general relativity within the density range modeled by the EOS. 
Instead, this phase transition takes place right above $n>n_\tov$. 
Therefore, the onset of this phase transition coinciding with $n_\tov$ is accidental. 
If one wants to study the possibility of a phase transition determining the onset of the unstable branch, phase transitions must be included in the EOS models to a significant fraction, as in Ref.~\cite{Gorda:2022lsk}.
For $I_\pQCD = 0.75$ a qualitatively similar softening is required but there is more freedom in the EOSs for $n> n_\term$.
By contrast, an EOS with $I_\pQCD = 0.5$ is largely unconstrained above $n > n_{\tov}$. 

In order to visualize the behavior above $n_{\term}$ explicitly, we have chosen representative EOSs that fulfill the QCD constraint at $n_{\term} = n_\tov$ but not necessarily above. 
For each of these EOSs, we generate 5000 extensions between $n_{\tov} < n < 15 n_{\sat}$ using the GP approach.
The extensions are then conditioned with the QCD constraint at $15 n_\sat$ so that the behavior between $n_{\tov} < n < 15 n_{\sat}$ remains consistent with the QCD input.  
These extensions show how the EOS must evolve beyond $n_\tov$ given its behavior below $n_\tov$. 
Figure~\ref{fig:extension_ensemble} shows the extensions for representative EOSs with $I_\pQCD = 0.5, 0.75$ and $0.85$ (for $I_\pQCD=0.98$, our prior does not contain any valid samples). 
We have checked that these EOSs are indeed representative of EOSs with these pQCD tension indices.
We find a difference in behavior above $n_\tov$ based on the value of the pQCD tension index at the TOV point $I_\pQCD(n_\tov)$.
For the highest index of $I_\pQCD(n_\tov) = 0.85$ shown in Fig.~\ref{fig:extension_ensemble}, we observe that the EOS must soften  drastically above $n_{\tov}$, with $c_s^2(n) \leq 0.1$ all the way to  $n = 15 n_\sat$. 
Furthermore, for this EOS, the value of the index at $15 n_\sat$ remains in the range $I_\pQCD(15 n_\sat) \in [0.9, 1]$, indicating that the EOSs needs to remain very soft.
For the smaller value of the pQCD tension index $I_\pQCD(n_\tov)=0.75$, we similarly observe that the index at $15 n_\sat$ remains close to its value at $n_\tov$ and the EOS must continue to remain soft.
Finally, for the lowest $I_\pQCD(n_\tov)= 0.5$ shown, the EOS extensions can obtain significantly different values of the pQCD tension index at $15 n_\sat$.

We quantify the degree of softening as a function of pQCD tension index in Fig.~\ref{fig:change_of_index}, where for 100 EOSs in our posterior with fixed values of the pQCD tension index $I_\pQCD(n_\tov)$, we show the distribution of the average speed of sound of the extensions on the interval $n \in [8, 15] n_\sat$.
From this figure we see that the average $c_s^2$ for the extensions of EOSs with $I_\pQCD = 0.85, 0.75$ are strongly skewed to small values, below $c_s^2 \lesssim 0.03$ and $c_s^2  \lesssim 0.11$, respectively (95\% credence).
By contrast, the extensions of EOSs with $I_\pQCD = 0.5$ have a very large spread in the average $c_s^2$, with a shape that follows the prior distribution, indicating that there is no forced behavior in this case.

\begin{figure*}
\centering
  \includegraphics[height=0.25 \textwidth]{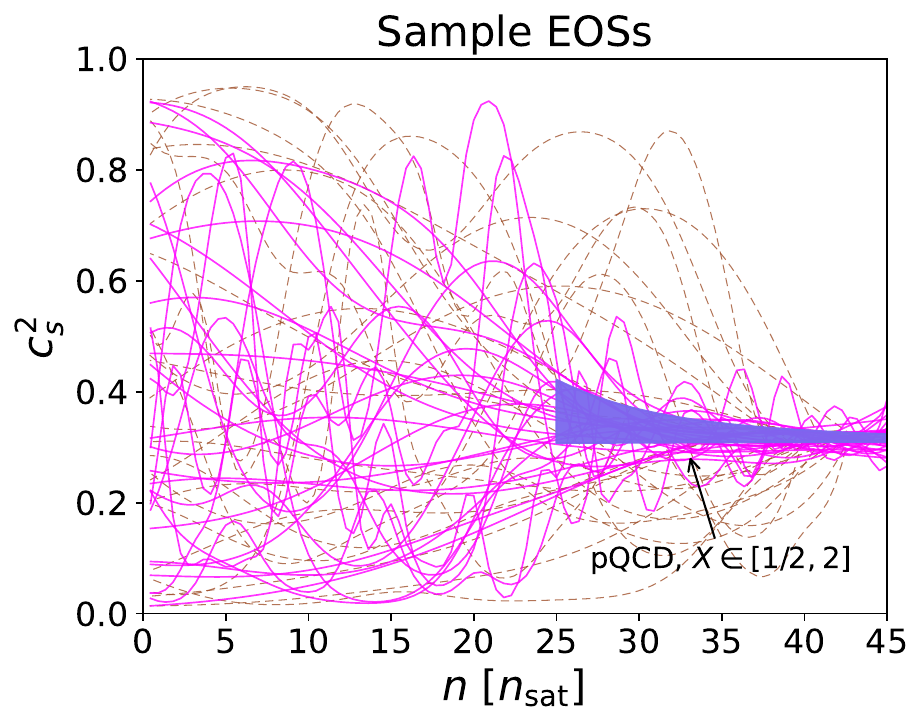}
  \includegraphics[height=0.26 \textwidth]{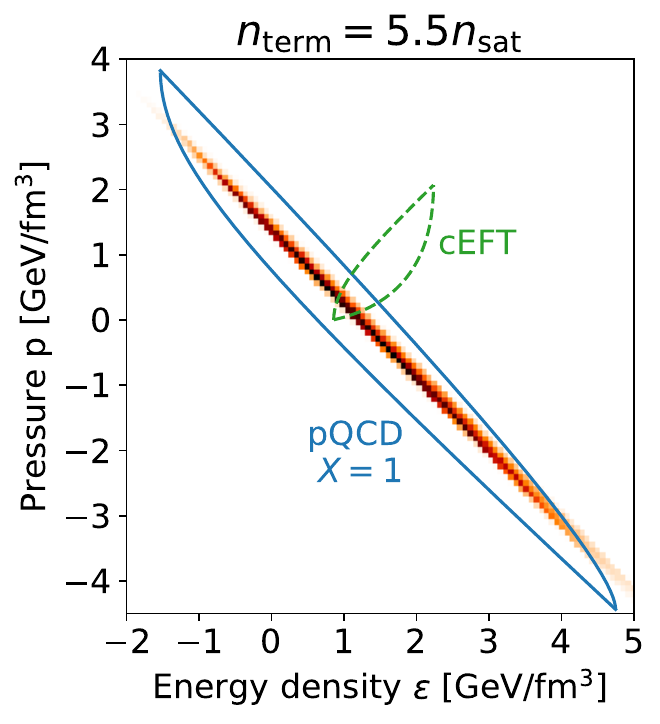} 
  \includegraphics[height=0.25 \textwidth]{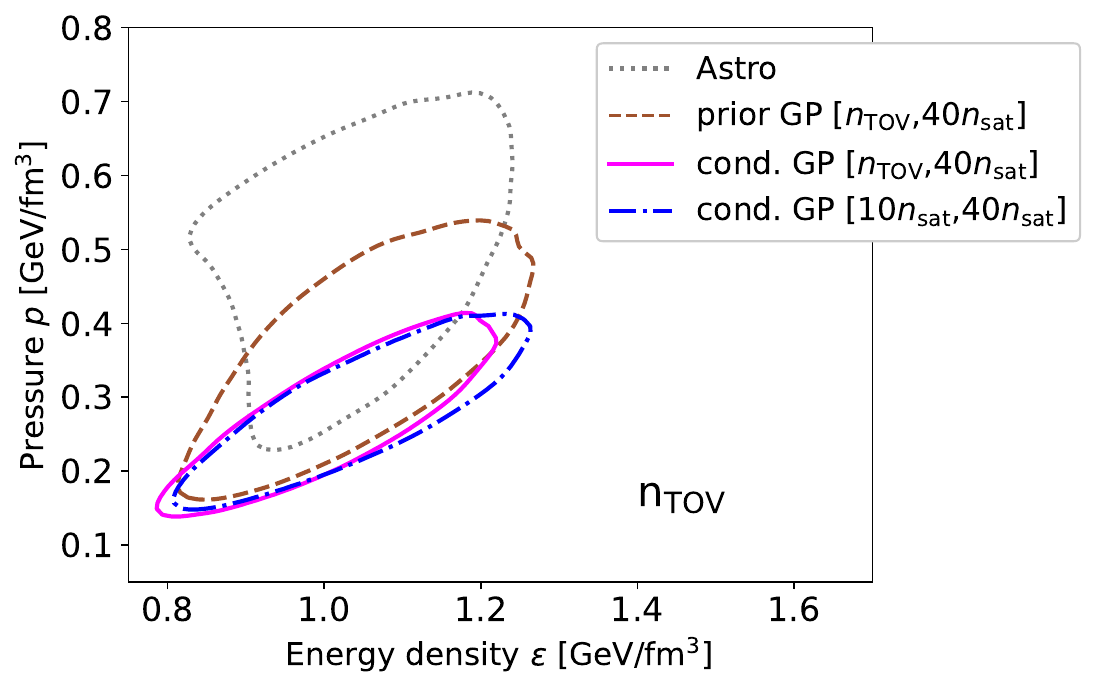} 
\caption{(\textbf{Left}) A sample of EOSs drawn from the hierarchical models of EOS extensions with correlation length $\ell \sim \mathcal{U}(1n_\sat,20n_\sat)$, the width $\eta \sim \mathcal{N}(1.25, 0.25^2)$ and average $\bar{c}_s^2 = \mathcal{N}(0.3,0.3^2)$. 
Negative values of $\eta$ and $\bar{c}_s^2$ are removed. 
The dashed prior GP is anchored to the pQCD EOS at $\mu_\pQCD = 2.6$~GeV, while the solid magenta GP is conditioned with the pQCD speed of sound on the interval $n \in [25, 40] n_\sat$ (see main text). 
(\textbf{Middle}) The resulting QCD likelihood function at $n_\term = 5.5 n_\sat$ for the GP conditioned on the pQCD $c_s^2(n)$ for $n \in [25, 40]n_\sat$. 
This termination density corresponds approximately to the mean value of $n_\tov$ for our ensembles after applying the astrophysical input. 
Shown in blue and green are the constraints arising from the chiral EFT calculation of Ref.~\cite{Hebeler:2013nza} and the $X = 1$ pQCD calculations of Ref.~\cite{Gorda:2021znl} alone, using the construction of Ref.~\cite{Komoltsev:2021jzg}. 
(\textbf{Right}) The posterior  $\varepsilon$--$p$ regions at $n_\tov$ after marginalizing over the EOS extensions between $n_\term$ and 40$n_\sat$.}
\label{fig:extensions_hierarchical}
\end{figure*}

In the previous sections, we have seen that the constraining power of the QCD input increases as function of the termination density when going from $n_\term = n_\tov$ to $n_\term = 1.2 n_\tov$; see, e.g., Fig.~\ref{fig:likelihood-panels}.
In Fig.~\ref{fig:tesion} we show the distribution of the pQCD tension indices $I_\pQCD(n_\tov)$ for EOSs that are allowed at $n_\term = 1.2 n_\tov$ and for EOS that are allowed at $n_\term = n_\tov$ but rejected before $n_\term = 1.2 n_\tov$.
The majority of the EOS that are excluded at higher $n_\term$ are EOSs which have large indices $I_\pQCD(n_\tov)$, forcing the EOS to undergo a rapid softening beyond $n_\tov$. 
This behavior cannot be excluded by the rigorous physics principles behind our implementation of the QCD constraint, but it would be interesting to understand what microphysical behavior could lead to such behavior. 

\subsection{\texorpdfstring{Marginalization over EOS extensions beyond $n_{\term}$}{Marginalization over EOS extensions beyond n\_term}}
\label{sec:marginalization}

The termination of the explicit EOS models at $n_\term$ puts the EOS below and above $n_\term$ in an asymmetric role; below the termination density our prior contains a large number of non-extreme EOSs by construction. 
Above $n_\term$, even the most extreme EOS extension of Fig.~\ref{fig:extension_cartoon} is realized without any penalty.
This has been criticized in Ref.~\cite{Essick:2023fso}, in which the authors point out that the construction of Ref.~\cite{Komoltsev:2021jzg} is similar to performing a maximization, rather than a marginalization over the likelihoods of different possible extensions.
As our final point in this work, we address the marginalization over a set of extensions of the EOS beyond $n_\term$. 
We do so by constructing a model of possible high-density EOSs down from the pQCD EOS at $\mu_\pQCD = 2.6$~GeV, distinct from the prior model of NS EOSs below $n_\term$.
Having an explicit model for EOSs also above $n_\term$ allows us to incorporate further theoretical inputs. 
Since the pQCD speed of sound is well converged at high densities, see Fig.~\ref{fig:extension_cartoon}, we can further condition this model with the pQCD speed of sound. 
We may do so either only at the pQCD matching point $n_\pQCD \approx 40 n_\sat$(referred to below as ``prior''), or over a larger range of densities where the speed of sound is under good theoretical control, $[25, 40]n_\sat$ (``conditioned''). 

For modeling the EOS above $n_\term$, we define another hierarchical GP model anchored to the (scale-averaged) pQCD EOS defined with the following hyperparameters:
\begin{equation}
\begin{split}
\ell& \sim \mathcal{U}(1n_\sat, 20n_\sat)\,, \\
\eta& \sim \mathcal{N}(1.25, 0.25^2)\,, \\
\bar{c}_s^2& \sim \mathcal{N}(0.3, 0.3^2)\, ,
\end{split}
\end{equation}
see Fig.~\ref{fig:extensions_hierarchical} (left).
Because of the large interval in density between the typical $n_\term$ and $n_\mathrm{QCD} \approx 40n_\sat$, we choose a more general prior for the correlation length $\ell$ here than was done in Ref.~\cite{Gorda:2022jvk}. 
We also choose the prior on $\bar{c}_s^2$ to have a mean centered near the conformal value of $1/3$, but allow it to vary broadly within the model.
To be conservative with the conditioned GP, we take the standard deviation of the GP at a given $n \in [25, 40]n_\sat$ to equal twice of that of the scale-averaged pQCD calculation.

The procedure for applying this high-density prior as a model for EOS extensions is as follows. 
First, we use either the low-density GP or CSM to model the NS EOS up to a specified termination density $n_\term$.
Then, for this $n_\term$, we calculate a QCD likelihood function by constructing the posterior distribution for different $\varepsilon$--$p$ values that arise from marginalizing over the new high-density GP model above $n_\term$. 
Concretely, we calculate the distribution by performing a kernel-density estimation of the $\varepsilon$--$p$ values of the high-density GP model at $n_\term$. 
This posterior distribution is then reinterpreted as a new QCD likelihood function within the NS EOS inference at $n_\term$ that marginalizes over the hierarchical model above, as opposed to the construction from Ref.~\cite{Komoltsev:2021jzg}, which instead returns a weight of 1 for any $\varepsilon$--$p$ point within the allowed region at $n_\term$ for a fixed $X$.
More details of this procedure are given in Appendix~\ref{sec:marginalization_details}.

In Fig.~\ref{fig:extensions_hierarchical} we show in the left panel a small sample of EOSs drawn from these models of EOS extensions, showing a broad range of possible behaviors. 
The conditioning of the magenta GP is clearly visible. 
In the middle panel we show the QCD likelihood function resulting from the conditioned GP as applied at {$n_\term = 5.5n_\sat$}, compared to the constraints arising from Ref.~\cite{Komoltsev:2021jzg} for the chiral EFT and pQCD constraints. 
This value of $n_\term$ approximately corresponds to the mean of the $n_\tov$ for our ensembles using only the astrophysical input (see Fig.~\ref{fig:likelihood-panels}). 
We see that the likelihood function is mostly peaked in the interior of the $X = 1$ pQCD region. 
As has been pointed out already in Ref.~\cite{Gorda:2022lsk}, the impact of the QCD input within the inference of the NS EOS occurs primarily along the upper boundary of the pQCD region in Fig.~\ref{fig:extensions_hierarchical}, since the astrophysical constraints already reduce the allowed $\varepsilon$--$p$ region in the neighborhood of the other extreme boundaries (cf.~also the discussion of the upper edge of the $\varepsilon$--$p$ regions in Fig.~\ref{fig:e-p_panel1}). 

Turning to the rightmost panel of Fig.~\ref{fig:extensions_hierarchical}, we see the effect of the marginalization over the GP model of EOS extensions directly as the change in the allowed $\varepsilon$--$p$ region at $n_\tov$. 
As expected from our observations of the central panel, we see a shift towards smaller $p$ values in this figure. 
The prior GP shows a similar shift to the effect of using a termination density that is about 20-40\% larger without the marginalization (cf.~the middle and right columns of Fig.~\ref{fig:e-p_panel1}), while the conditioned GP shows a much more pronounced softening, more comparable to the effect of the QCD input applied at $n_\term = 10n_\sat$ (cf.~Fig.~\ref{fig:e-p_panel1}).
In both cases, the quantitative effect is to shift the posterior pressure distribution within the most massive NSs to smaller values, with the conditioned GP shifted more strongly. 
That the upper limit for the conditioned GP is significantly below that of the prior demonstrates that the additional constraint is indeed due to the conditioning and not the assumed prior. 
This decreased upper bound is natural, as the pQCD sound speed used in the conditioning is inconsistent with the stiff high-density segment seen in Fig.~\ref{fig:extension_cartoon}, thus affecting the EOS at lower densities as well.

Lastly, we observe that the credible regions of $\varepsilon$--$p$ are almost indistinguishable between $n_\term = n_\tov$ and $10 n_\sat$, demonstrating independence of this parameter when using the conditioned GP extension. 

\section{Discussion and conclusions}
\label{sec:discussion_and_conclusions}

\begin{figure*}
\centering
\includegraphics[width = 0.43\textwidth]{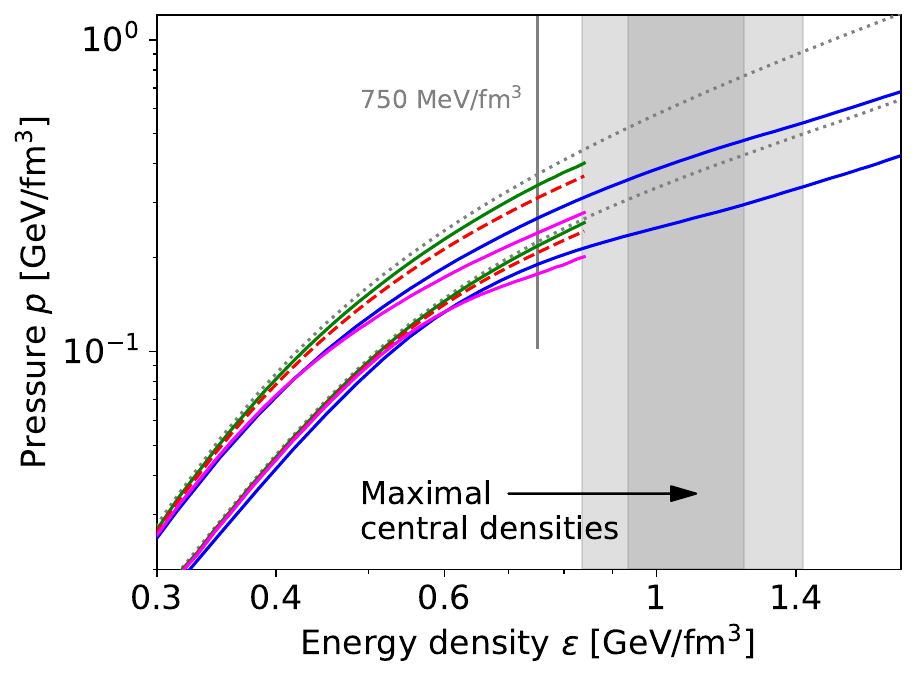}
\includegraphics[width = 0.56\textwidth]{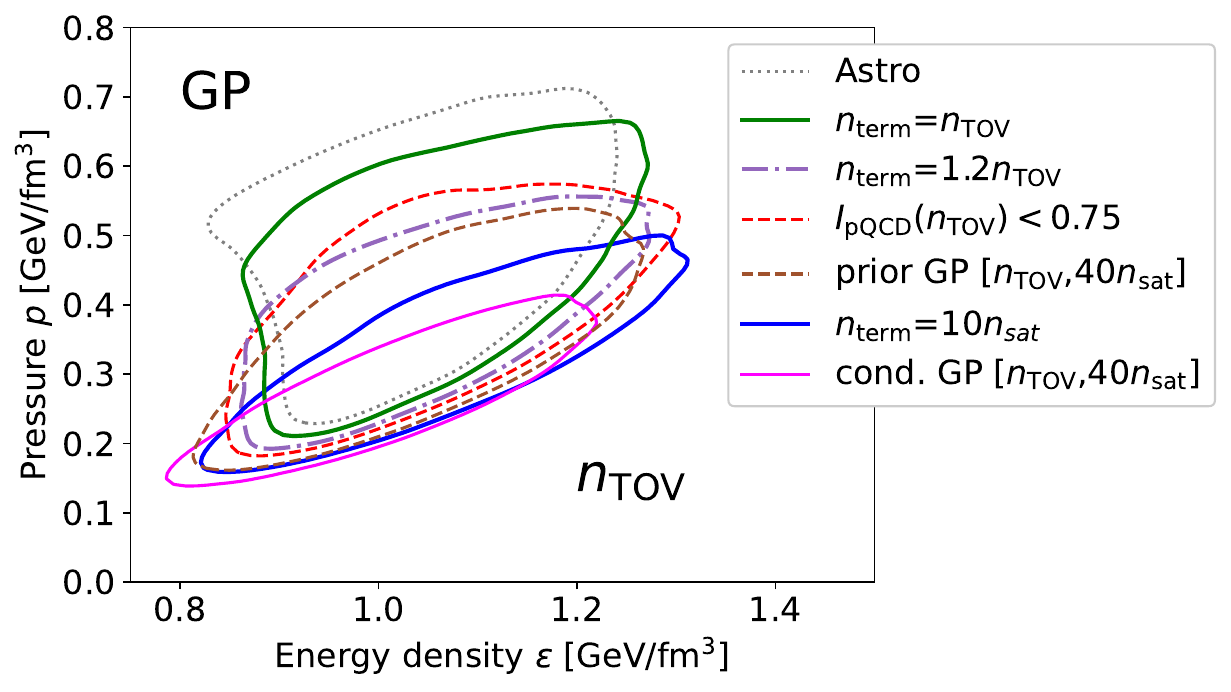}
\caption{The impact of QCD input at the densities reached in the cores of maximally massive NS using different prescription for the EOS extension. 
(\textbf{Left}) Pressure as a function of energy density. For those prescriptions with $n_\term = n_\tov$ the 68\% credible intervals for the pressure are extended only up to lower bound of the 95\% credible interval of the TOV energy densities. 
Above this point we cannot access the posterior distribution $P( p | \varepsilon )$ but have access only to the conditional distribution  $P( p | \varepsilon , \varepsilon < \varepsilon_\tov)$. (\textbf{Right}) The posterior  $\varepsilon$--$p$ regions at $n_\tov$.}
\label{fig:softening}
\end{figure*}

In this paper, we have studied the impact of the high-density QCD input on the modeling of the dense-matter EOS, and explained the different findings presented in the literature. 
In particular, using the most general QCD input we found that the constraining power of the QCD input depends sensitively on the chosen termination density $n_\term$. 
For the conservative choice of $n_{\term}=n_{\tov}$, the impact of the QCD input is largely removed, while the constraining power rises fast with $n_{\term}$. 
This difference in constraining power just beyond $n_\tov$ arises from EOSs that undergo a change in behavior that is qualitatively different from the assumed priors.
In particular, we have shown that most EOSs that are removed for $n_{\term}>n_{\tov}$ have a large pQCD tension index, leading to an abrupt softening above $n_{\tov}$ that is disfavored when the EOS priors employed here are constructed to higher densities.
Hence, to remain in agreement with the QCD input, the EOS needs either to soften below $n_{\tov}$ or it has to exhibit a drastic softening immediately above $n_{\tov}$ that persists for an extended density range before stiffening again dramatically, see also Ref.~\cite{Annala:2019puf}.
This behavior is seen in the EOS extensions depicted in Fig.~\ref{fig:extension_ensemble}.
We reiterate that the strong softening coinciding with $n_\tov$ is accidental, as $n_\tov$ is already set by the behavior of the EOS at lower densities.
While it cannot be excluded by the generic thermodynamic considerations we employ, it would be interesting to see what microphysical behavior would be necessary to realize such behavior within dense matter.
Note, however, that high speeds of sound above $n \gtrsim 25 n_\sat$ are inconsistent with pQCD calculations.

We note that different choices for $n_\term$,  $n_\term = n_\tov$ and  $n_\term > n_\tov$, have their merits and answer slightly different questions: the choice of the termination density eventually reflects the phenomenological question at hand. 
On the one hand, the EOS is a fundamental prediction of QCD and exists independently of NSs at densities above $n_{\tov}$. 
The EOS above $n_{\tov}$ is required in modeling the post-merger phase of binary NS mergers (see Refs.~\cite{Ujevic:2023vmo, Tootle:2022pvd}%
\footnote{Note that Ref.~\cite{Ujevic:2023vmo} found that multi-messenger signals from binary NS merges are not effective at constraining the EOS above $n_\tov$, rendering the QCD input the only source of information in this density range in the near future.})
and sets a prediction for the quantity that may be verified in the future using, e.g., lattice field theory\footnote{On this point, see in particular two recent works in Refs.~\cite{Moore:2023glb,Fujimoto:2023unl}.}.
In order to address these questions, the EOS must be modeled beyond the densities reached in stable NSs, and  $n_\term > n_{\tov}$ is required. 
On the other hand, one can be more conservative  and use $n_\term = n_\tov$ if one chooses to infer only the EOS along the stable NS branch.
This choice avoids prior-dependent modeling of the unstable branch, allowing for more robust conclusions.
Similarly, for an even more conservative modeling of NSs, one could imagine terminating the EOS model at $n_\term = n(2.1 M_\odot )$ which is the mass of the heaviest observed neutron star.
Whatever application one has in mind, it is crucial that the QCD input in the form of Ref.~\cite{Gorda:2022jvk} be applied at the highest density that will be studied. 
It is inconsistent to apply the QCD input at $n_\term$ and then consider the model for even higher densities $n>n_\term$.

Since the pQCD speed of sound is well converged to even lower densities than the pressure~\cite{Gorda:2021znl,Gorda:2023mkk}, we have additionally constructed a QCD likelihood function that incorporates this additional information. 
We have done so by constructing a further GP model down from the pQCD EOS calculation and conditioning it with the pQCD calculation of $c_s^2$. 
This also addresses the points raised in Ref.~\cite{Essick:2023fso}, as this new likelihood function involves marginalizing over the new GP above $n_\term$.
We find that the conditioned GP gives results that are independent of $n_\term \geq n_\tov$ (see Fig.~\ref{fig:extensions_hierarchical}) and agrees with results obtained for $n_\term = 10 n_\sat$, see Fig.~\ref{fig:softening}. 
The conditioned likelihood function is made publicly available in Ref.~\cite{komoltsev_2024_10592568}. 
We also see in Fig.~\ref{fig:softening} that the unconditioned GP leads to similar results as choosing $n_\term = 1.2n_\tov$ or by restricting the pQCD tension index $I_\pQCD(n_\tov) < 0.75$.

Finally, we discuss the limitations of our study.
An important finding of this work is that the QCD input removes EOS that are very stiff around $n_{\tov}$. 
However, we allow for extreme stiffening of the EOS at low densities, around $1-2 n_{\sat}$, where EFT approaches can otherwise provide constraints on the EOS and typically suggest the EOS to be soft~\cite{Tews:2018kmu,Drischler:2020hwi}.
In this work, we have not varied the maximal density up to which we consider EFT input, which might bias our EOS sets to be stiffer than if we marginalized over the maximal EFT density~\cite{Capano:2019eae,Essick:2020flb}. 
We also have not marginalized over the scale $\mu_\pQCD$ where we evaluate the pQCD values; marginalizing over lower values than used here would provide a more restrictive QCD input~\cite{Gorda:2023usm}.
Furthermore, we only considered two EOS priors that model the EOS in terms of the speed of sound and are, hence, rather similar.
We have also not considered astrophysical data from other X-ray sources such as quiescent low-mass X-ray binaries or photospheric-radius--expansion burst, which decrease the upper bound on the radii with respect to the NICER X-ray data~\cite{Al-Mamun:2020vzu}.  
Finally, for the purpose of marginalizing over EOS extensions above $n_\term$, it would be interesting to investigate other prior choices for the EOS model, such as GP models with a density dependent correlation length.

\acknowledgements
We thank Reed Essick, Joonas Hirvonen, D\'ebora Mroczek, Jorge Noronha, Jaki Noronha-Hostler, Krishna Rajagopal, and Rachel Steinhorst for useful discussions.
This work has been supported in part by the Deutsche Forschungsgemeinschaft (DFG, German Research Foundation) project-ID 279384907--SFB 1245, by the State of Hesse within the Research Cluster ELEMENTS (projectID 500/10.006), and by the ERC Advanced Grant ``JETSET: Launching, propagation and emission of relativistic jets from binary mergers and across mass scales'' (Grant No. 884631) (T.G.).
R.S. acknowledges support from the Nuclear Physics from Multi-Messenger Mergers (NP3M) Focused Research Hub which is funded by the National Science Foundation under Grant Number 21-16686, and by the Laboratory Directed Research and Development program of Los Alamos National Laboratory under project number 20220541ECR.
J.M. is supported by the CNRS/IN2P3 NewMAC project, and are also grateful to PHAROS COST Action MP16214 and to the LABEX Lyon Institute of Origins (ANR-10-LABX-0066) of the \textsl{Universit\'e de Lyon} for its financial support within the program \textsl{Investissements d'Avenir} (ANR-11-IDEX-0007) of the French government operated by the National Research Agency (ANR).
The work of I.T. was supported by the U.S. Department of Energy, Office of Science, Office of Nuclear Physics, under contract No.~DE-AC52-06NA25396, by the Laboratory Directed Research and Development program of Los Alamos National Laboratory under project number 20230315ER, and by the U.S. Department of Energy, Office of Science, Office of Advanced Scientific Computing Research, Scientific Discovery through Advanced Computing (SciDAC) NUCLEI program.

Author contributions: O.K. performed most of the numerical analysis, with contributions from R.S. 
T.G. wrote the scripts for the marginalization and conditioning of the high-density GP. 
All authors contributed to the preparation and revision of the manuscript.

\appendix
\section{Marginalization above $n_{\tov}$}
\label{sec:marginalization_details}

In this appendix we present some details related to the marginalization procedure discussed in Sec.~\ref{sec:marginalization} and elaborate on why using the likelihood function discussed in the section corresponds to marginalization over the EOS prior above $n_{\rm TOV}$.

We start by having an EOS prior in the whole density region between the low density and pQCD limits, described by some prior probability distribution $P(\eta_l , \eta_h)$ where the collective variables $\eta_l$ and $\eta_h$ correspond to the EOS (or more specifically speed of sound) below and above an arbitrary density $n_\term$. 

In computing a posterior distribution of a physical quantity $Q$ depending on the EOS, we are to marginalize over the EOS prior
\begin{equation}
P(Q) = \int d\eta_l d \eta_h P(Q|\eta_l, \eta_h) P(\eta_l, \eta_h)
\end{equation}

Now lets make a simplifying assumption that the prior below and above $n_\term$ is not correlated (or is weakly correlated) such that
\begin{equation}
P(\eta_l, \eta_h)  = P(\eta_l)P( \eta_h)
\end{equation}
This is a good approximation beyond the correlation length of the GP. 

Now, if the quantity $Q$ does not depend on the high-density EOS $\eta_h$ in any other way than through the end point $p_L$ and $\epsilon_L$ at $n_\term$
\begin{equation}
P(Q|\eta_l, \eta_h) = P(Q|\eta_l; p_L, \epsilon_L),
\end{equation}
as is the case in stable NSs if $n_\term = n_\tov$, we may perform the integral over the $\eta_h$ independent of the observable $Q$ by marginalizing over $p_L$ and $\epsilon_L$
\begin{align}
P(Q) & = \int d\eta_l d \eta_h P(Q|\eta_l, \eta_h) P(\eta_l)P(\eta_h) \\
& = \int d\eta_l d \eta_h dp_L d\epsilon_L P(Q|\eta_l; p_L, \epsilon_L)P(\eta_l) \\
& \quad \quad \times  P(p_L, \epsilon_L| \eta_h) P(\eta_h)\\
& = \int d\eta_l dp_L d\epsilon_L P(Q|\eta_l; p_L, \epsilon_L)P(\eta_l) w(p_L, \epsilon_L),
\end{align}
where
\begin{equation}
w(p_L, \epsilon_L) \equiv \int d \eta_h P(p_L, \epsilon_L| \eta_h) P(\eta_h)
\end{equation}
is the QCD likelihood function plotted in the center panel of Fig.~\ref{fig:extensions_hierarchical}.

\bibliography{refs.bib}

\end{document}